\documentclass[runningheads]{llncs}
\usepackage{amsmath}
\usepackage{graphicx}
\usepackage{hyperref}
\begin{document}
\title{Cranial Implant Prediction using Low-Resolution 3D Shape Completion and High-Resolution 2D Refinement}
\titlerunning{Abbreviated paper title}
%
\author{Amirhossein Bayat\inst{1,2,3}\and
Suprosanna Shit\inst{1,3} \and Adrian Kilian\inst{4} \and Jürgen T. Liechtenstein\inst{4} \and Jan S. Kirschke\inst{2,3} \and
Bjoern H. Menze\inst{1,3}}
\authorrunning{A. Bayat et al.}
\institute{
Department of Informatics, Technical University of Munich, Germany \and
Department of Neuroradiology, Klinikum rechts der Isar, Germany \and
TranslaTUM Center for Translational Cancer Research, Munich, Germany \and
Department for Oral \& Maxillofacial Surgery, University Hospital Schleswig-Holstein, Campus Kiel, Arnold-Heller-Strasse 3, 24105 Kiel, Germany
\email{amir.bayat@tum.de}}

%
\maketitle              
\begin{abstract}
Designing of a cranial implant needs a 3D understanding of the complete skull shape. Thus, taking a 2D approach is sub-optimal, since a 2D model lacks a holistic 3D view of both the defective and healthy skulls. Further, loading the whole 3D skull shapes at its original image resolution is not feasible in commonly available GPUs. To mitigate these issues, we propose a fully convolutional network composed of two subnetworks. The first subnetwork is designed to complete the shape of the downsampled defective skull. The second subnetwork upsamples the reconstructed shape slice-wise. We train both the 3D and 2D networks in tandem in an end-to-end fashion, with a hierarchical loss function. Our proposed solution accurately predicts a high-resolution 3D implant in the challenge test case in terms of dice-score and the Hausdorff distance.

\end{abstract}
\section{Introduction}
Cranial implant design is a crucial task for clinical planning of cranioplasty \cite{li2020online}. Previous works mainly rely on freely available CAD tools for cranial implant design \cite{Gallinproceedings,Chenarticle,Marzolaarticle,Janarticle}. The time requirements and need for expert intervention for these approaches are a major hindrance for fast and in-prompt deployment. The AutoImplant challenge aims to look for simple and easy-to-use automatic solution that can accurately predict cranial implants. Keeping this in mind, we tailor our proposed solution to best fit the requirements of clinicians.

Previous literature \cite{Angeloarticle} tend to exploit the geometric symmetry and predict cranial implant based on the unaffected skull region. Nevertheless, this results in a suboptimal solution, since the human skull is not perfectly symmetric in reality. These solutions also fall short when the implant is not exclusively in one hemisphere. Morais et al. \cite{Morais2019} used a deep 3D encoder-decoder \cite{bayat2020inferring,sekuboyina2020verse,fracture2019conditioned} network to reconstruct the incomplete skull in low-resolution space. While the low-resolution space facilitates faster processing, the quality of the reconstruction lacks minute local anatomical detail. In the Autoimplant baseline paper \cite{li2020baseline}, a similar approach is taken where the authors first localize the defective region in the skull and then predict the implant using an encoder-decoder network. While this pipeline is suitable for modular design of accurate defective region detection and implant prediction, the network is not end-to-end trainable, and thus any error during the first stages would penalize the implant prediction.

\begin{figure}[t!]

\centering
\includegraphics[width=0.24\linewidth,trim={0 0 0 80},clip]{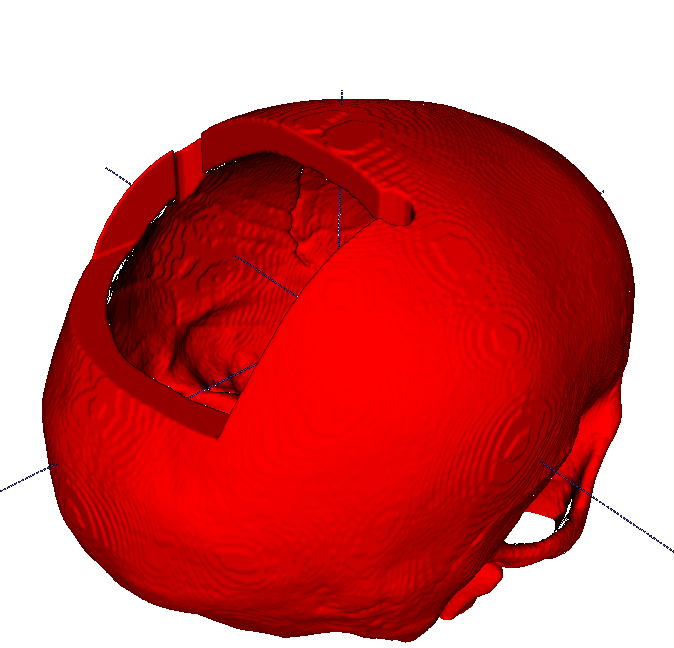}
\includegraphics[width=0.24\linewidth,trim={0 0 0 80},clip]{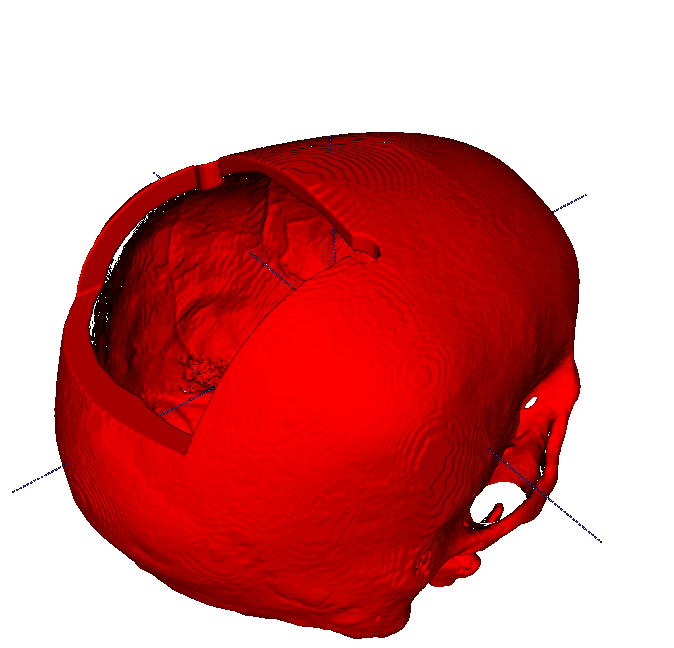}
\includegraphics[width=0.24\linewidth,trim={0 0 0 80},clip]{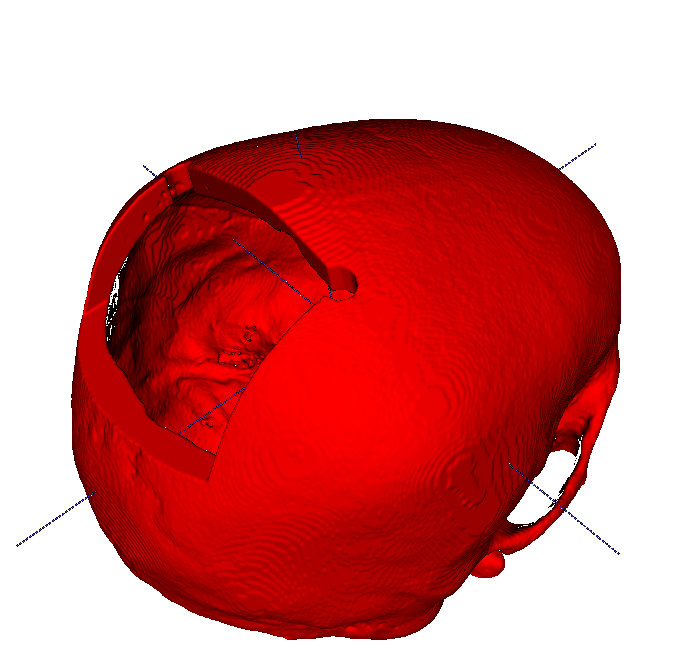}
\includegraphics[width=0.24\linewidth,trim={0 0 0 80},clip]{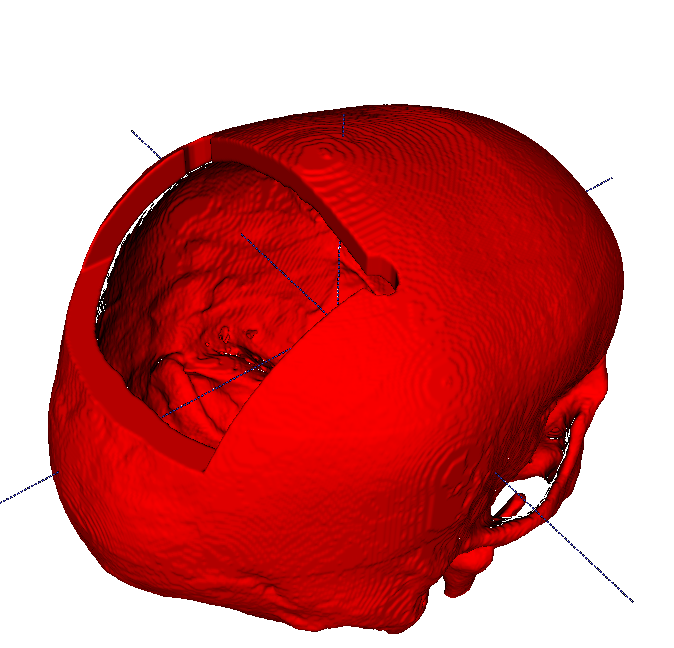}

\includegraphics[width=0.24\linewidth,trim={0 0 0 80},clip]{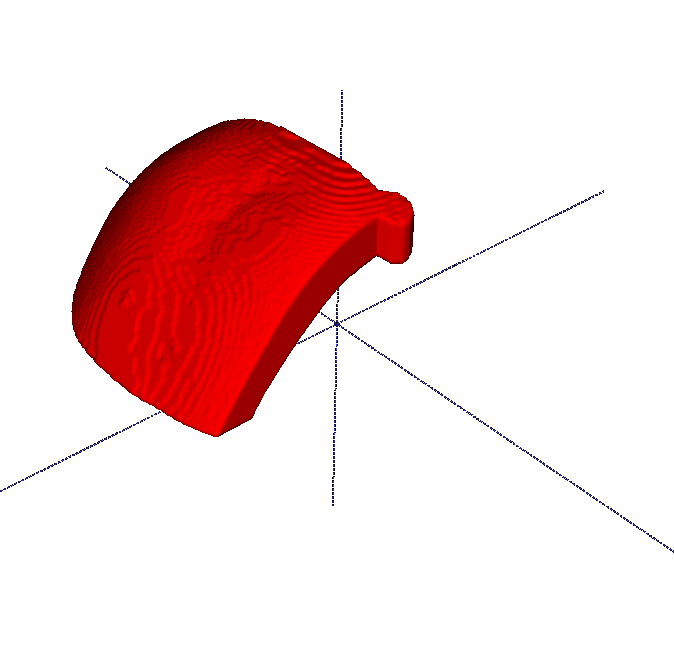}
\includegraphics[width=0.24\linewidth,trim={0 0 0 80},clip]{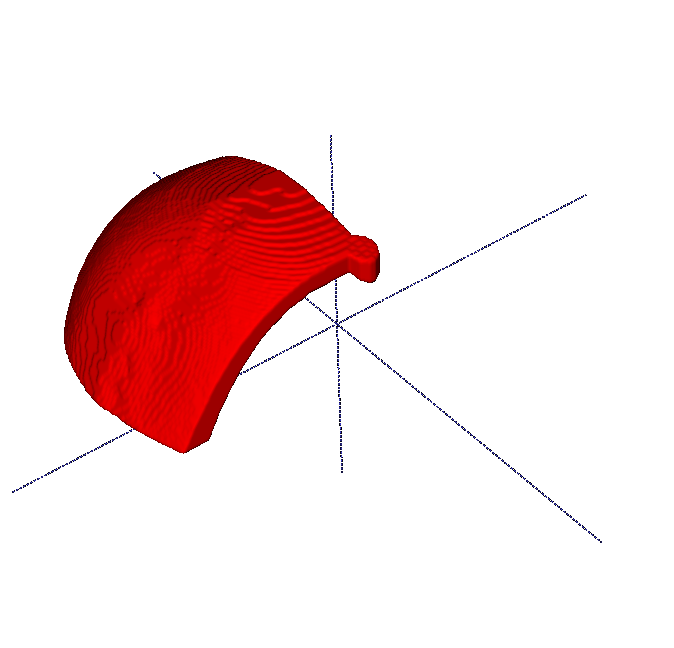}
\includegraphics[width=0.24\linewidth,trim={0 0 0 80},clip]{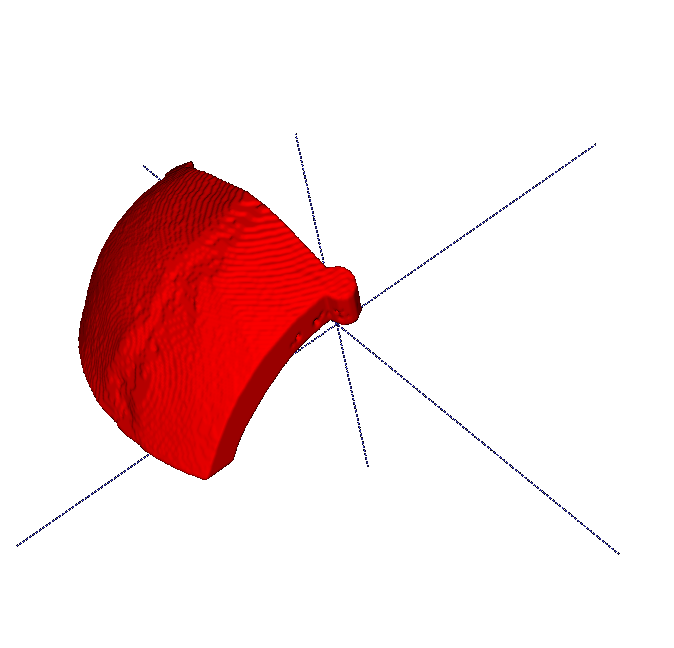}
\includegraphics[width=0.24\linewidth,trim={0 0 0 80},clip]{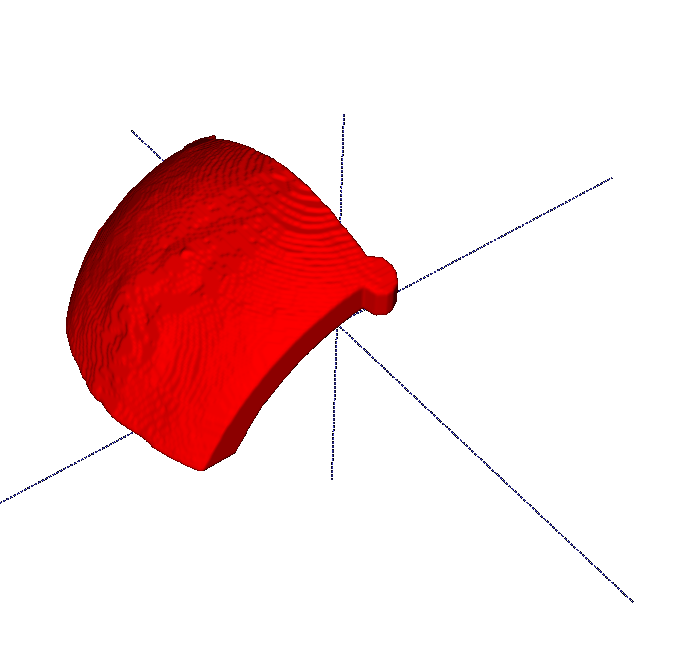}

\caption{\footnotesize \textbf{Few Training sample:} The first row depicts rendered 3D volumes of four randomly selected defective scan from the training dataset. The second row shows the corresponding ground truth cranial implant.}
\label{train_example}
\end{figure}

In this approach, we try to alleviate this by relying on coarse-scale implant prediction in 3D followed by fine-scale enhancement of the predicted implant. We identify that 3D is most suitable to predict the implant since anatomical consistency is best captured by a 3D receptive field compared to any local 2D slices. However, to reduce the memory and computational power, we first predict the implant in a down-sampled defective skull. Subsequently, we enhance the predicted implant slice wise by a 2D decoder network. Thus our solution becomes end-to-end trainable and also is efficient at the same time for the high-resolution implant prediction task.
\section{Method}
The dataset is created by artificially generating the defect in the scan \cite{li2020baseline}. Thus the original skull would be the ground truth for the implant prediction task. We leverage this availability of the target label and cast the implant prediction task as a supervised volumetric reconstruction task. At the core of our method lies a 3D encoder-decoder network. This network takes the low-resolution defective skull as input and predicts a low-resolution implant at the output. We argue that the implant prediction task lies in a lower-dimensional manifold since the key properties to predict implant are the inner and outer surface consistency. Hence, a down-sampled input space is sufficient for a coarse-scale identification of the implant region. A simple element-wise subtraction of the reconstructed skull and the input will produce the desired implant. This approach is in line with the shape completion literature \cite{Dai2016ShapeCU,Sung2015DatadrivenSP,Sarmad2019RLGANNetAR,Stutz2018Learning3S,Han2017HighResolutionSC}. Next, we need to upsample the predicted implant, which can be done in several ways. Classical approaches, such as spline-based interpolation, can be a simple choice. Alternatively, a decoder network proved to be superior in the super-resolution task \cite{hu2020feedback,bhowmik2017training}. Hence, we incorporate a second module in our method, a 2D up-sampler. This up-sampler takes selected axial slices during training and predicts the up-sampled version of it. To be able to train the both the network jointly and also fit the data in the GPU memory, we select $N$ random slices out of the reconstructed shape and select the corresponding slices from the original scale Ground Truth. The error between the predicted slice and the ground truth skull is used to train the 2D decoder. The high-resolution reconstruction error, along with the 3D shape completion error, contributes to the training of the 3D encoder-decoder.

\begin{figure*}[t!]
\label{met}
\centering
\includegraphics[width=1.0\textwidth]{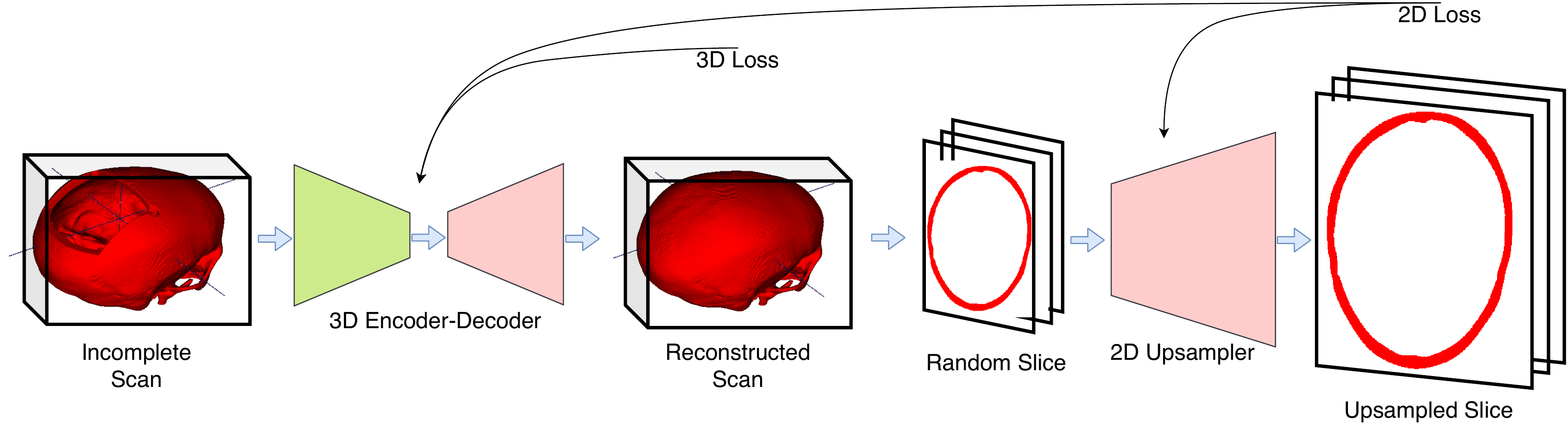}

\caption{\footnotesize \textbf{Schematic overview of our proposed pipeline for predicting the cranial implant.} The downsampled defective scan goes through an encoder-decoder based shape completion network. During training, $N$ number of random reconstructed skull goes through a second decoder network for high-resolution reconstruction. For the 3D shape completion, we use a volumetric $\ell_1$ norm, and for the 2D refinement task, we use summation of 2D $\ell_1$ loss.}
\end{figure*}

\subsection{Network Architecture \& Loss Function}

In the following, we describe the architecture of two subnetworks in our model and the loss functions used to train the model.

\subsubsection{3D Encoder-Decoder:} Encoder-decoder type network has been previously used in bio-physical simulation \cite{ezhov2020real}, image segmentation \cite{navarro2019shape} etc. Our 3D network has three sequential components, such as an encoder, bottleneck, and a decoder. The encoder further compresses the input signals into a more compact representation, which is processed in the bottleneck unit to extract useful features. These features go through the decoder part to reconstruct the complete skull. The complete architecture is as follows:\\
$IN_1\rightarrow CN^1_{64}\rightarrow CN^2_{64} \rightarrow CN^2_{64} \rightarrow RB_{64} \rightarrow RB_{64} \rightarrow RB_{64} \rightarrow RB_{64} \rightarrow TC^2_{64} \rightarrow TC^2_{64} \rightarrow C^1_{1}\rightarrow OUT_1$

\noindent where $IN_1$ and $OUT_1$ is input and output volume respectively with single channel, $CN^s_{\#ch}$ is convolution with stride $s$ and output channel $\#ch$ followed by batch norm and ReLU, $TC^s_{\#ch}$ is transposed convolution with stride $s$ and output channel $\#ch$ followed by batch norm and ReLU, $RB_{\#ch}$ is residual block consists of two successive unit of convolution with stride 1 and output channel $\#ch$ followed by instance norm and ReLU, and $C^s_{\#ch}$ is convolution with stride $s$ and output channel $\#ch$ followed by sigmoid. Note that all convolution and norm layers described here are 3D.

\subsubsection{2D Decoder Upsampler:} The 2D upsampler network consists of four residual blocks, followed by the nearest neighborhood upsampling layer and a final convolution layer. The residual blocks refine the low-resolution reconstructed scans to incorporate anatomical consistency, which aids precise high-resolution skull at the output. We concatenate the corresponding slice of the defective scan along with the reconstructed scan and pass it as an input to the 2D upsampler. This helps to correct any location-wise mismatch in the 3D shape-completion task. Borrowing a few notations defined in the previous paragraph, the complete architecture is given below:\\
$IN_2\rightarrow CN^1_{64}\rightarrow SE_{64}\rightarrow RB_{64}\rightarrow SE_{64}\rightarrow RB_{64}\rightarrow SE_{64}\rightarrow RB_{64}\rightarrow SE_{64}\rightarrow RB_{64} \rightarrow NN^{sqrt(512/180)}_{64} \rightarrow NN^{sqrt(512/180)}_{64} \rightarrow C^{1}_{1}\rightarrow OUT_1$\\
where $SE{\#ch}$ is `squeeze and excitation' layer and $NN^s_{\#ch}$ is Nearest Neighborhood (NN) upsample with scale factor $s$ and output channel $\#ch$ followed by instance norm and ReLU. Note that all convolution and norm layers described here are 2D.

\subsubsection{Loss Function:}
Let's denote the ground truth data at original scale as $I_G$, downsampled ground truth data $I_g$, defective 3D volume at original scale as $I_D$, downsampled defective 3D volume as $I_d$, the functional form of the 3D encoder-decoder network as $S()$, and the functional form of the 2D upsampler network as $U()$ respectively. The cranial implant is predicted as follows:
\begin{equation}
    \mbox{Cranial Implant }= U(S(I_d))\setminus I_D
\end{equation}

\noindent where $\setminus$ denotes set difference. The total loss function of our method is as follows:
\begin{align}
\mathcal{L}_{total} & = \mathcal{L}_{3D} + \mathcal{L}_{2D}\\
\mathcal{L}_{3D} & =\|S(I_d) -I_g\|_{\ell_1} \\ 
\mathcal{L}_{2D} & = \sum_{i\in\Omega}\| U(S(I_d)^i) - I_G^i\|_{\ell_1}
\end{align}
where $\Omega~\mbox{is the set of random slices} $
\subsection{Implementation}
We realize our model in PyTorch. We trained the networks with Adam optimizer and a learning rate of 0.0001. We used an Nvidia Quadro P6000 GPU. The batch size for the 3D network was 1, so one volume per iteration. \textbf{We downsampled the original 3D volume by a factor of $\frac{512}{180}$ in all dimension because that is the largest 3D volume we can fit in our GPU along with the 2D decoder module. The downsampled 3D volume is zero-padded in the z-dimension to make it $180\times180\times180$.} After predicting the completed 3D shape in low resolution, we sample 10 slices randomly along the Z-axis and concatenate them channel-wise with the downsampled corresponding slice from the defective skull and feed them to the upsampler decoder. We can't fit the entire volume with the original scale in the memory, so we have to select 2D slices. In order to avoid overfitting, we select the slices randomly. It is important to note that, after downsampling the volume with a $\frac{512}{180}$ scaling factor, every 3 slices along Z-axis in the original scale correspond to 1 slice in the downsampled volume. Thus, after reconstructing the 3D shape, we have a set of selected slices using {random indices} and three sets of {[random indices/0.35]}, {[random indices/0.35]}+1 and {[random indices/0.35]}+2.  We select the corresponding slices from the defective scan and downsample them in 2D to be concatenated with the slices from the predicted shape. Thus, the batch size for the upsampler decoder network is 30.

\subsubsection{Inference:}

For inference, similarly, a downsampled volume is fed to the network, and it is reconstructed in low resolution using the first sub-network. After that, all of the slices along the Z-axis are fed to the upsampler decoder, one-by-one, and stacked in volume to reconstruct the shape in 3D. Subsequently, we subtract the defective input scan from the high-resolution reconstructed scans to estimate the cranial implant. Finally, as a post-processing step, we erode and dilate the segmentation consequently with a sphere structure with a radius of 2 to remove the noise. Subsequently, we select the largest component in the segmentation map, using connected component analysis.

\section{Experimental Results}
\begin{table}[h]
    \centering
    \footnotesize
    \caption{\label{tab:score} Our score on the validation dataset }
    \setlength\tabcolsep{4pt} 
    \renewcommand{\arraystretch}{1.5} 
    \begin{tabular}{|l|c|c|}
        \hline
        Method & \textbf{Dice} & \textbf{HD-distance} \\ \hline
        Ours (Transposed Conv) & 0.8363 & 10.6570 \\ \hline
        Ours (NN Upsampling) & \textbf{0.9358} & \textbf{7.6100} \\ \hline
    \end{tabular}
    
\end{table}

We work with $100$ data samples split $5:1$ forming the training and validation set. We validate our approach by comparing the constructed implants to the ground truth, using the Dice score and the Hausdorff distance. The validation results are presented in Table \ref{tab:score}. \textbf{We experimented with two variations of our 2D decoder model. In the first case, we trained the original decoder, and in the second case, we replace the NN upsampling layers of the 2D decoder with transposed convolution layers.} We observe that the 2D decoder with NN upsampling layers performs significantly better than the decoder with transposed-convolution layers. We attribute this to the over parameterization during the upsampling step. Since the image is binary in nature, the nearest neighborhood upsampling layer is sufficient for this task.

\begin{figure}[t!]

\centering
\includegraphics[width=0.24\linewidth,trim={200 250 200 180},clip]{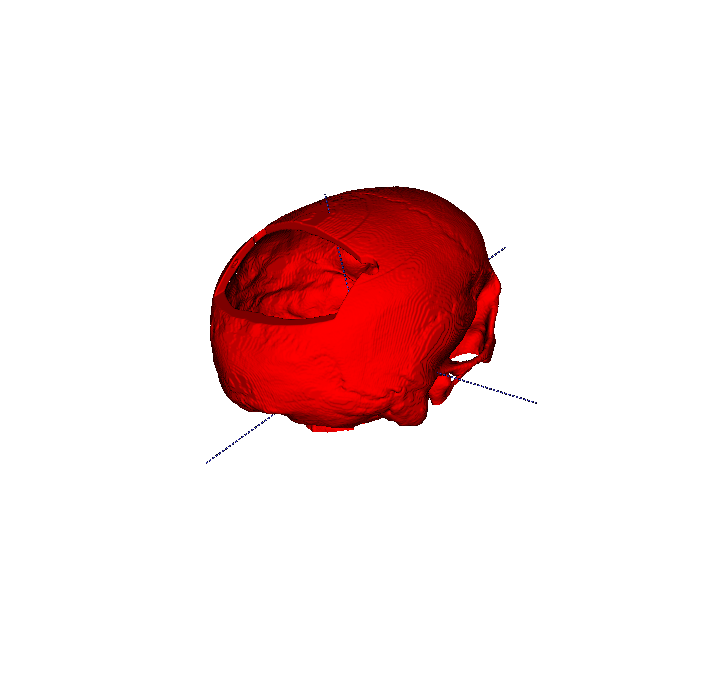}
\includegraphics[width=0.24\linewidth,trim={200 200 200 180},clip]{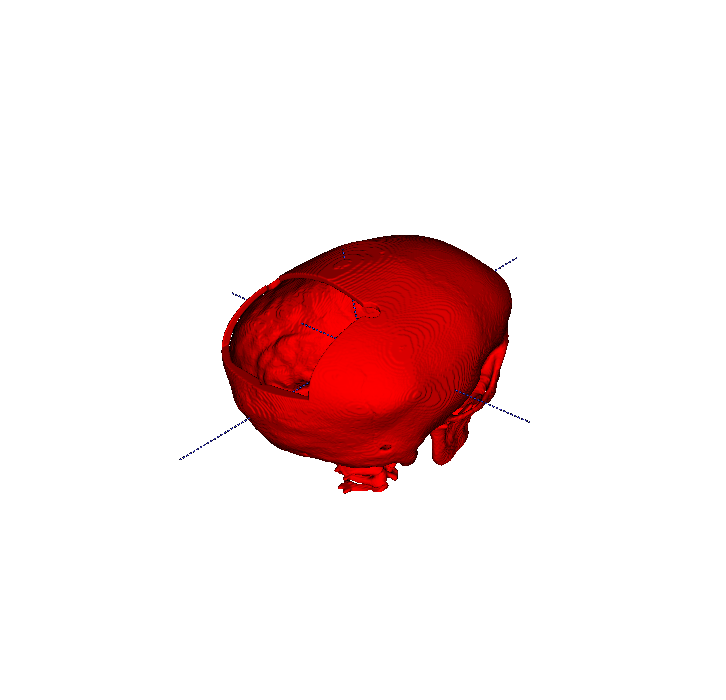}
\includegraphics[width=0.24\linewidth,trim={200 200 200 180},clip]{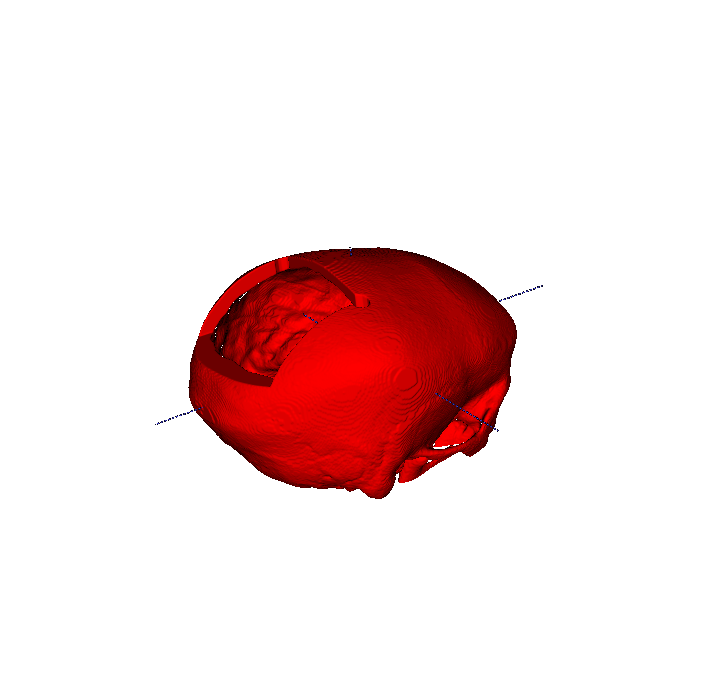}
\includegraphics[width=0.24\linewidth,trim={200 200 200 180},clip]{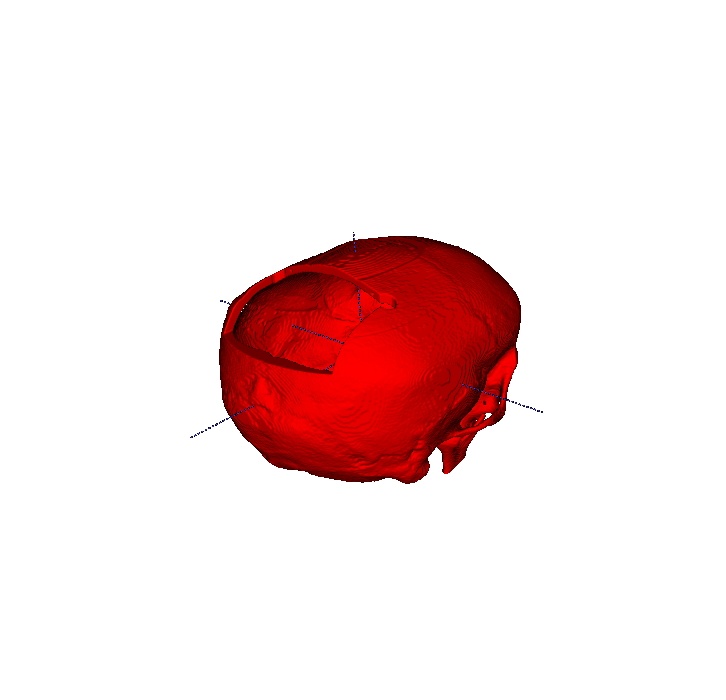}

\includegraphics[width=0.24\linewidth,trim={200 250 200 180},clip]{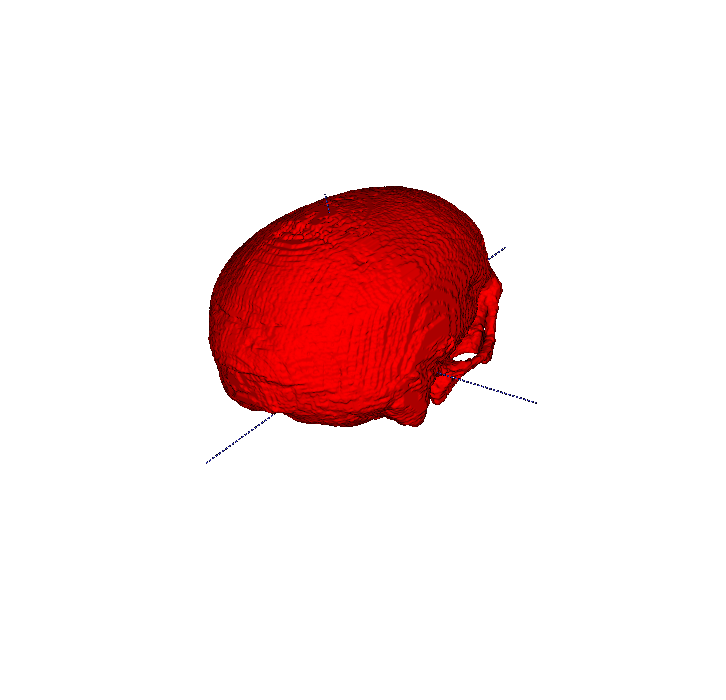}
\includegraphics[width=0.24\linewidth,trim={200 200 200 180},clip]{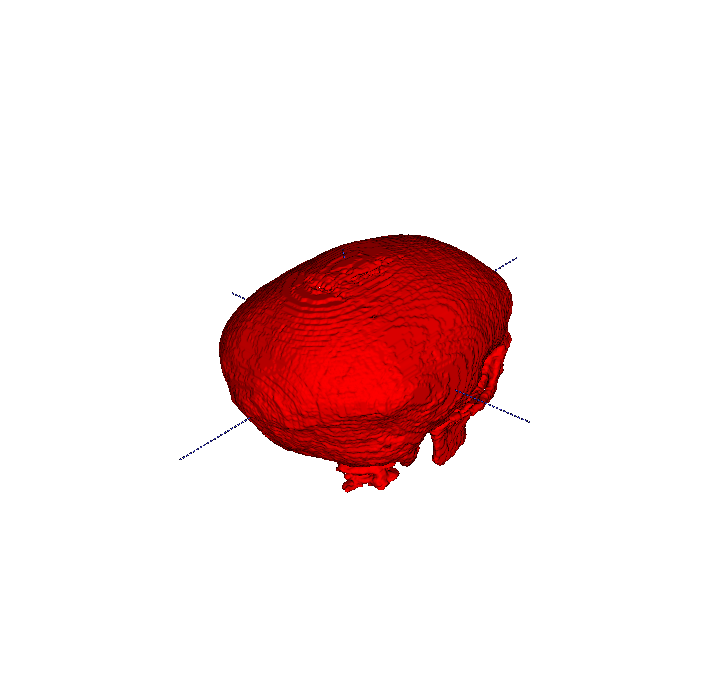}
\includegraphics[width=0.24\linewidth,trim={200 200 200 180},clip]{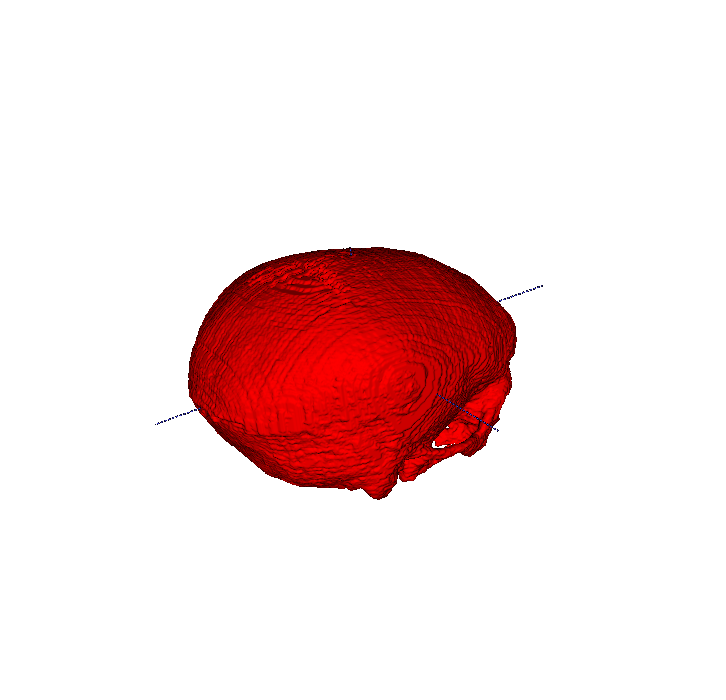}
\includegraphics[width=0.24\linewidth,trim={200 200 200 180},clip]{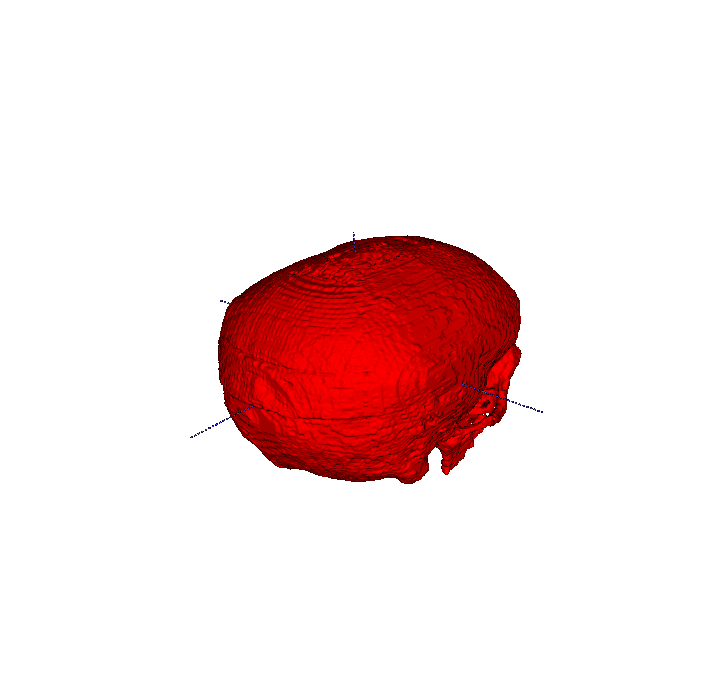}

\includegraphics[width=0.24\linewidth,trim={200 250 200 180},clip]{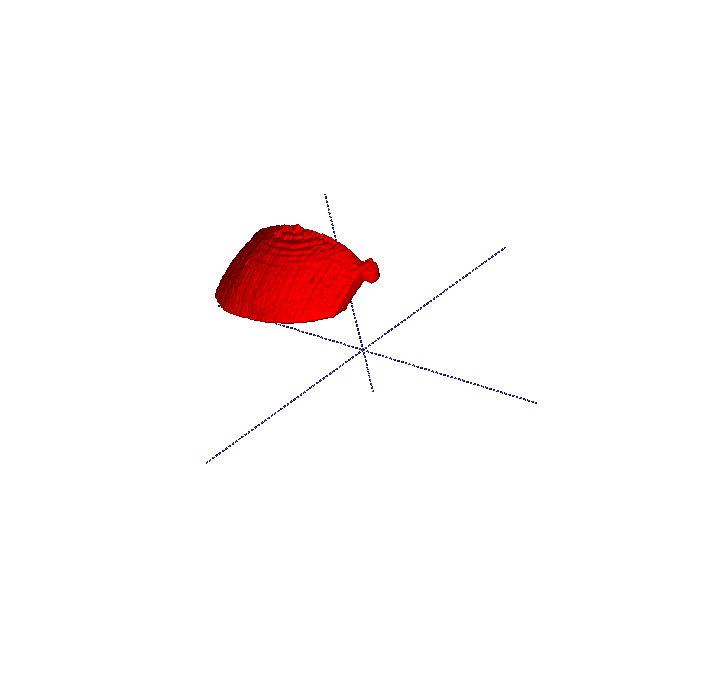}
\includegraphics[width=0.24\linewidth,trim={200 200 200 180},clip]{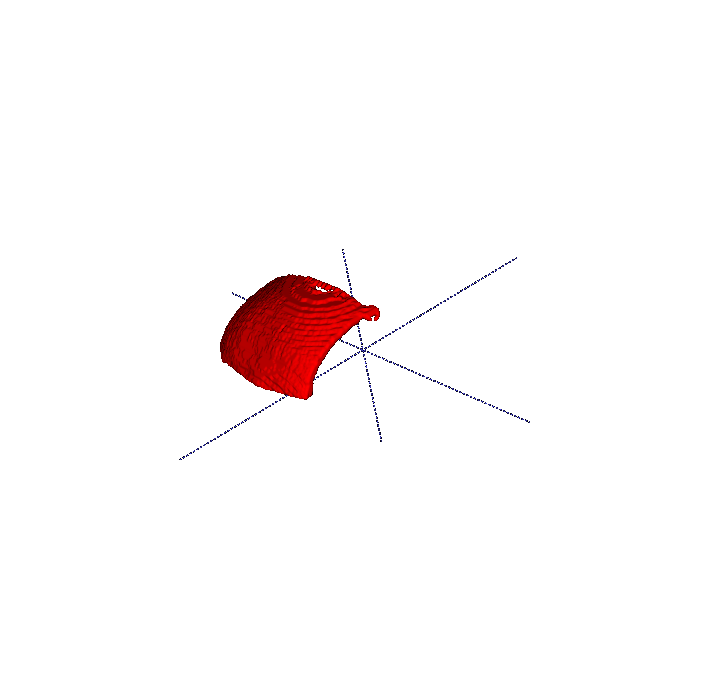}
\includegraphics[width=0.24\linewidth,trim={200 200 200 180},clip]{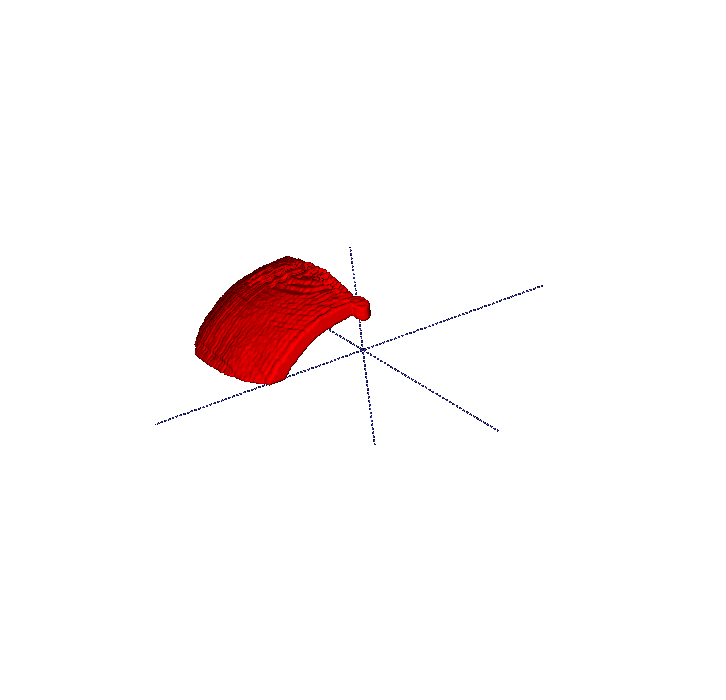}
\includegraphics[width=0.24\linewidth,trim={200 200 200 180},clip]{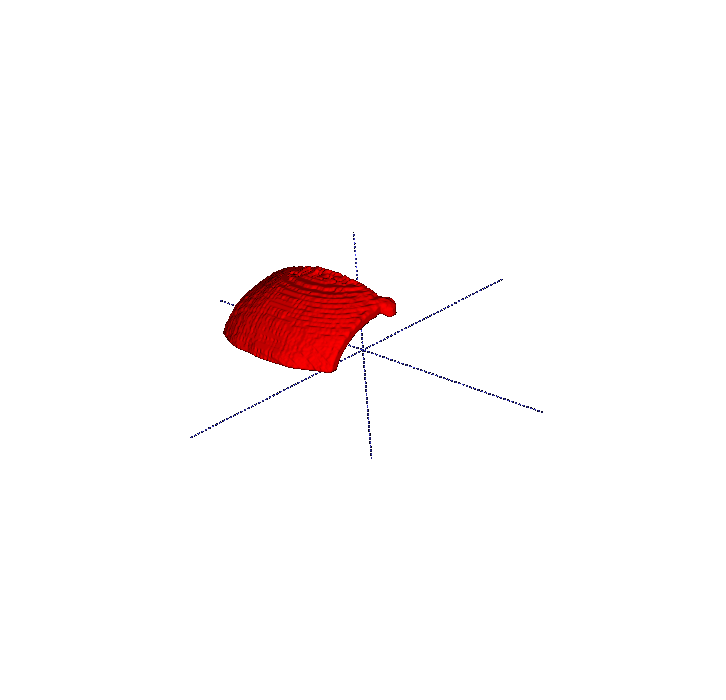}

\caption{\footnotesize \textbf{Qualitative results:} The first row depicts four rendered 3D volume of defective scan from the test dataset. The second row shows the reconstructed skull by our method of the corresponding defective skull. The third row is the corresponding cranial implant predicted by our method. We observe that our method generalizes well and accurately reconstruct the skull to predict the cranial implants.}
\label{aneu_result}
\end{figure}

\begin{table}[h]
    \centering
    \caption{\label{tab:test} Our score on the 100 test cases.}
    \setlength\tabcolsep{4pt} 
    \renewcommand{\arraystretch}{1.5} 
    \begin{tabular}{|l|c|c|}
        \hline
        Method & \textbf{Dice} & \textbf{HD-distance} \\ \hline
        Baseline \cite{li2020baseline} & 0.8555 & 5.1825 \\ \hline
        Ours (NN Upsampling) & \textbf{0.8957} & \textbf{4.6019} \\ \hline
    \end{tabular}
\end{table}

Finally, we tested our model on the challenge test set and report the results for cases  $000\sim099$ in Table \ref{tab:test}. Our model outperforms the baseline method proposed in \cite{li2020baseline} both in the Dice score and Hausdorff distance. We did not report the results on cases $100\sim109$, since the location of the defect is very different from the training set, and the model could not predict the implant. Fig. \ref{aneu_result} shows the qualitative results of randomly selected scans from the test data set. Visual inspection also confirms that our model estimates accurate cranial implants for these cases.
The source code of our model is accessible from \href{https://github.com/mlentwicklung/autoimplant}{https://github.com/mlentwicklung/autoimplant}.





%
%
%
%
%
\section{Conclusion}
We provide an efficient and compact solution for the AutoImplant 2020 challenge, which is suitable for fast and easy deployment. Our key innovation is the incorporation of a two-stage reconstruction policy, where the first stage predicts a coarse-scale implant, and the second stage super-resolve it to a high-resolution one. We achieve accurate implant prediction on the validation dataset. Our model is end-to-end in the high-resolution space and thus can serve as a baseline for developing more complex models aiming to better learn the anatomically invariant implant prediction.

\section*{Acknowledgement}
Amirhossein Bayat is supported by the European Research Council (ERC) under the European Union's `Horizon  2020' research \& innovation programme (GA637164–iBack–ERC–2014–STG). Suprosanna Shit is supported by the Translational Brain Imaging Training Network (TRABIT) under the European Union's `Horizon 2020' research \& innovation program (Grant agreement ID: 765148).

{ \bibliographystyle{splncs04}
\bibliography{mybibliography}
}

\end{document}